\documentstyle[twoside,12pt,epsf]{article}
\let\ssection=\section
\renewcommand{\section}{\setcounter{equation}{0}\ssection}

\def\and{
{\LARGE\em \&}
\bigskip
}

\def\foot{
\markboth{}{}
\cleardoublepage}

\def \half {\raise1pt\hbox{$\scriptstyle
        {1 \over 2}\displaystyle$}}
\def \twentyfour {\raise1pt\hbox{$\scriptstyle
       {1 \over {24}}\displaystyle$}}
\def \sixth {\raise1pt\hbox{$\scriptstyle
       {1 \over 6}\displaystyle$}}

\catcode`\@=11

\def\ov{\overline}
\def\wh{\widehat}
\def\wt{\widetilde}
\def\be{\begin{equation}}
\def\ee{\end{equation}}
\def\ba{\begin{eqnarray}}
\def\ea{\end{eqnarray}}
\def\hg {{\cal HG}_{x_0}}
\def\hgp {{\cal HG}_p}
\def\agb {\overline {{\cal A}/{\cal G}}}
\def\agpb {\overline{{\cal A}/{\cal G}_p}}
\def\agp {{\cal A}/{\cal G}_p}
\def\ag {{\cal A}/{\cal G}}

\def\l{\lambda}

\def\s{\sigma}

\def\Lapop{\displaystyle{{\hbox to 0pt{$\sqcup$\hss}}\sqcap}}

\def\R{{\rm I\!R}}
\def\N{{\rm I\!N}}

\def\Q{{\mathchoice
{\setbox0=\hbox{$\displaystyle\rm Q$}\hbox{\raise 0.15\ht0\hbox to0pt
{\kern0.4\wd0\vrule height0.8\ht0\hss}\box0}}
{\setbox0=\hbox{$\textstyle\rm Q$}\hbox{\raise 0.15\ht0\hbox to0pt
{\kern0.4\wd0\vrule height0.8\ht0\hss}\box0}}
{\setbox0=\hbox{$\scriptstyle\rm Q$}\hbox{\raise 0.15\ht0\hbox to0pt
{\kern0.4\wd0\vrule height0.7\ht0\hss}\box0}}
{\setbox0=\hbox{$\scriptscriptstyle\rm Q$}\hbox{\raise 0.15\ht0\hbox to0pt
{\kern0.4\wd0\vrule height0.7\ht0\hss}\box0}}}}
\def\C{{\mathchoice
{\setbox0=\hbox{$\displaystyle\rm C$}\hbox{\hbox to0pt
{\kern0.4\wd0\vrule height0.9\ht0\hss}\box0}}
{\setbox0=\hbox{$\textstyle\rm C$}\hbox{\hbox to0pt
{\kern0.4\wd0\vrule height0.9\ht0\hss}\box0}}
{\setbox0=\hbox{$\scriptstyle\rm C$}\hbox{\hbox to0pt
{\kern0.4\wd0\vrule height0.9\ht0\hss}\box0}}
{\setbox0=\hbox{$\scriptscriptstyle\rm C$}\hbox{\hbox to0pt
{\kern0.4\wd0\vrule height0.9\ht0\hss}\box0}}}}

\font\fivesans=cmss10 at 4.61pt
\font\sevensans=cmss10 at 6.81pt
\font\tensans=cmss10
\newfam\sansfam
\textfont\sansfam=\tensans\scriptfont\sansfam=\sevensans\scriptscriptfont
\sansfam=\fivesans
\def\sans{\fam\sansfam\tensans}
\def\Z{{\mathchoice
{\hbox{$\sans\textstyle Z\kern-0.4em Z$}}
{\hbox{$\sans\textstyle Z\kern-0.4em Z$}}
{\hbox{$\sans\scriptstyle Z\kern-0.3em Z$}}
{\hbox{$\sans\scriptscriptstyle Z\kern-0.2em Z$}}}}
\def\semi{\bigcirc\kern-1em{s}\;}

\newcount\foot
\def\note#1{\footnote{${}^{\number\foot}$}{\ftn #1}\advance\foot by 1}

\def\frac#1#2{{#1\over #2}}
\def\text#1{\quad{\hbox{#1}}\quad}

\font\ftn=cmr8 scaled\magstephalf

\font\it=cmti10 scaled\magstephalf
\font\bf=cmbx10 scaled\magstephalf

\def\sh{{\cal SH}}
\def\shb{\overline{\sh}}

\def\hg{{\cal H}}
\def\S{{\cal S}}

\begin{document}
\cleardoublepage
\pagestyle{myheadings}

\title{
A Manifestly Gauge-Invariant Approach to Quantum Theories of
Gauge Fields}
\author{Abhay Ashtekar\thanks{Center for Gravitational Physics and
Geometry, Physics Department, Penn State University, University Park,
PA 16802-6300, USA. Supported in part by the NSF Grant PHY93-96246
and the Eberly research fund of Penn State University}
{}~~~~~
Jerzy Lewandowski\thanks{Institute of Theoretical Physics,
Warsaw University, ul Ho\.za 69, 00-681 Warsaw, Poland. Supported partially
by the Polish Committee for Scientific Research (KBN) through grant no.\
2 0430 9101}
{}~~~~~
Donald Marolf$^\ast$\\
Jos\'e Mour\~ao
\thanks{Dept. of Mathematics and Statistics, University of Cyprus, P.O.
Box 537, Nicosia, Cyprus. On leave of absence from Dept. F\'{\i}sica,
Inst. Sup. T\'{e}cnico, 1096 Lisboa, Portugal. }
{}~~~~~
Thomas Thiemann$^\ast$}
\date{July 1994}
\maketitle
\markboth{Ashtekar, Lewandowski, Marolf, Mour\~ao and Thiemann}
{Gauge-Invariant Approach to Quantum Gauge Theories}

\pagenumbering{arabic}

\begin{abstract}

In gauge theories, physical histories are represented by space-time
connections modulo gauge transformations. The space of histories is
thus intrinsically non-linear. The standard framework of constructive
quantum field theory has to be extended to face these {\it kinematical}
non-linearities squarely. We first present a pedagogical account of
this problem and then suggest an avenue for its resolution.

\end{abstract}

\section{Introduction}

As is well-known, for over 40 years, quantum field theory has remained
in a somewhat peculiar situation. On the one hand, perturbative
treatments of realistic field theories in four space-time dimensions
have been available for a long time and their predictions are in
excellent agreement with experiments.  It is clear therefore that
there is something ``essentially right'' about these theories. On the
other hand, their mathematical status continues to be dubious in all
cases (with interactions), including QED. In particular, it is
generally believed that the perturbation series one encounters here
can be at best asymptotic. However, it is not clear what exactly they
are asymptotic to.

This overall situation is in striking contrast with, for example,
non-relativistic quantum mechanics.  There, we know well at the outset
what the Hilbert space of states is and what the observables are. In
physically interesting models, we can generally construct the
Hamiltonian operator and show that it is self-adjoint.  We take
recourse to perturbation theory mainly to calculate its eigenvalues
and eigenvectors. Therefore, if the perturbation series turns out not
to be convergent but only asymptotic, we know what it is asymptotic
to. The theories exist in their own right and perturbative methods
serve as approximation techniques to extract answers to physically
interesting questions. In realistic quantum field theories, we do not
yet know if there is an underlying, mathematically meaningful
framework whose predictions are mirrored in the perturbative answers.
When one comes to QCD, the problem is even more severe. Now, it is
obvious from observations that the physically relevant phase of the
theory is the one in which quarks and gluons are confined. And this
phase lies beyond the grasp of standard perturbative treatments.

To improve this situation, the program of constructive quantum field
theory was initiated in the early seventies.  This approach has had
remarkable success in certain 2 and 3 dimensional models.  From a
theoretical physics perspective, the underlying ideas may be
summarized roughly as follows. Consider, for definiteness, a scalar
field theory. The key step then is that of giving meaning to the
Euclidean functional integrals by defining a rigorous version $d\mu$
of the heuristic measure ``$[\exp(-S(\phi))]\prod_{x} d\phi(x)$'' on
the space of histories of the scalar field, where $S(\phi)$ denotes
the action governing the dynamics of the model. The appropriate space
of histories turns out to be the space ${\cal S}'$ of (tempered)
distributions on the Euclidean space-time and regular measures $d\mu$
on this space are in one to one correspondence with the so-called
generating functionals $\chi$, which are functionals on the Schwarz
space ${\cal S}$ of test functions satisfying certain rather simple
conditions.  (Recall that the tempered distributions are continuous
linear maps from the Schwarz space to complex numbers.)

Thus, the problem of defining a quantum scalar field theory can be
reduced to that of finding suitable measures $d\mu$ on ${\cal S}'$, or
equivalently, ``appropriate'' functionals $\chi$ on ${\cal S}$.
Furthermore, there is a succinct set of axioms --due to Osterwalder
and Schrader-- which spells out the conditions that $\chi$ must
satisfy for it to be ``appropriate,'' i.e., for it to lead to a
consistent quantum field theory which has an associated Hilbert space
of states, a Hamiltonian, a vacuum and an algebra of observables.
This strategy has led to the rigorous construction of a number of
interesting theories in 2 and 3 dimensions such as the $\lambda
\phi^4$ and the Yukawa models and to an understanding of their
relation to perturbative treatments. A more striking success of these
methods is that they have led to a rigorous construction of the
Gross-Neveau model in 3-dimensions which, being non-renormalizable,
fails to exist perturbatively in the conventional sense.

These successes are remarkable. However, we believe that the framework
has an important limitation: As in the heuristic, theoretical physics
treatments \cite{PR,RR}, it is based on the assumption that the theories
under consideration can be considered to be {\it kinematically linear}
\cite{GJ,VR,DI}. That is, even though dynamical non-linearities are
properly incorporated, it is assumed that the {\it space} ${\cal S}'$
{\it of histories is a vector space.} This assumption permeates the
whole framework. In particular, in their standard form, all of the key
Osterwalder-Schrader axioms use this property. Now, for unconstrained
systems --such as the $\lambda \phi^4$-model-- this assumption is not
restrictive. However, for constrained systems it is generally
violated. An outstanding example is provided by the Yang-Mills theory.
Now, because of the presence of constraints, the system has gauge
freedom and the space of physically distinct histories is provided by
${\cal A}/{\cal G}$, the space of connections on the Euclidean
space-time modulo gauge transformations.  In dimensions $d+1> 2$, this
is a highly non-linear space with complicated topology. Rather than
facing this non-linearity squarely, one often resorts to gauge fixing,
ignores global problems such as those associated with Gribov
ambiguities, obtains a linear space and proceeds to apply the standard
techniques of constructive quantum field theory. To an outsider
the disparity between the ``roughness'' with which ${\cal
A}/{\cal G}$ is steam-rollered into a linear space and the
sophistication with which functional analysis is then used to
construct measures seems rather striking. It is natural to ask if one
can not modify the general framework itself and tailor it to the
kinematical non-linearities of ${\cal A}/{\cal G}$.

The purpose of this contribution is to suggest an avenue towards this
goal. (For earlier work with the same motivation, see \cite{As}.)
We should emphasize, however, that ours is only an approach: we
will not be able to present a definitive generalization of the
Osterwalder-Schrader axioms. Furthermore, the key steps of our
framework are rather loose. However, it {\it is} tailored to facing the
kinematical non-linearities of gauge theories squarely from the very
beginning.

None of the authors are experts in constructive quantum field theory.
The main ideas behind this approach came rather from an attempt to
construct quantum general relativity non-perturbatively. Consequently,
certain notions from quantum gravity --such as diffeomorphism
invariance-- play a non-trivial role in the initial stages of our
constructions. This is a strength of the framework in the context of
diffeomorphism invariant gauge theories. Examples are: the Yang-Mills
theory in 2 dimensions which is invariant under area-preserving
diffeomorphisms and the Chern-Simons theories in 3 dimensions and the
Husain-Kucha\v{r} model \cite{HK} in 4 dimensions which are invariant
under all diffeomorphisms. In higher dimensional Yang-Mills theories,
on the other hand, the action does depend on the background Euclidean
or Minkowskian geometry; it is not diffeomorphism invariant. In the
general program sketched here, this geometry {\it is} used in
subsequent steps of our constructions. However, we expect that, in a
more complete and polished version, it would play an important role
from the beginning.  It is clear that considerable work is still
needed to make the framework tight and refined, tailored more closely
to quantum Yang-Mills fields. As the title of the contribution
indicates, our primary aim is only to suggest a new approach to the
problem, thereby initiating a re-examination of the appropriateness of
the ``standard'' methods beyond the kinematically linear theories.

This contribution is addressed to working theoretical --rather than
mathe-\\matical-- physicists. Therefore, the presentation will be
somewhat pedagogical. In particular, {\it we will not assume prior
knowledge of the methods of constructive quantum field theory.} We
begin in section 2 with a summary of the idea and techniques used in
this area and indicate in particular how the appropriate measure is
constructed for the $\lambda \phi^4$ model. As indicated above, this
discussion will make a strong use of the kinematical linearities of
models considered. In section 3, we turn to gauge theories and show
how certain recent advances in the development of calculus on the
space of connections modulo gauge transformations can be used to deal
with the intrinsic kinematical non-linearities.  In section 4, we
indicate how these techniques can be used for Yang-Mills theories.  In
particular, we will construct the 2-dimensional Euclidean Yang-Mills
theory and indicate its relation to the Hamiltonian framework.  In
section 5, we summarize the main results, point out some of the
strengths and the limitations of this approach and discuss directions
for further work.

\section{Kinematically linear theories}

Several of the ideas we will use in the construction of measures on
infinite dimensional non-linear spaces are similar (but not
identical!) to those used in the linear case. Therefore, in this
section we will recall from \cite{GJ,VR,GV,Y} some well known results
about measures on infinite dimensional linear spaces in the context of
constructive quantum field theory.

Let us consider an Euclidean scalar field theory on flat Euclidean
space-time $M = \R^{d+1}$. It is natural to choose as the space $\sh$
of {\it classical} histories of the theory the {\it linear} space of
suitably regular (say $C^2$ and rapidly decreasing at infinity) scalar
field configurations; $\sh = \{\phi(x)\}$. The dynamics of the theory
is determined by the action functional $S$ on $\sh$
\be \label{2.3}
S(\phi) = \int_{\R^{d+1}} \left( {1 \over 2}
\partial_\mu \phi(x) \partial^\mu \phi(x) + V(\phi(x)) \right)
d^{d+1}x \ ,
\ee
where $V(\phi(x))$ denotes the self-interaction potential (which is
assumed to be bounded from below). In the ``classical" theory
\footnote{ We used quotation marks in the word ``classical" because,
strictly speaking, the solutions of the Euclidean equations of motion
do not play a direct role in classical physics; they have physical
interpretation only in the semi-classical approximation.}
we are interested in {\it points} in $\sh$ (i.e.  particular
histories) that correspond to the extrema of $S$, i.e.  in solutions
of the (Euclidean) equation of motion
\be
\label{2.4} \Delta \phi - {\partial V \over\partial \phi} = 0 \ .
\ee
If $V$ is cubic or higher order in $\phi$ the equation (\ref{2.4}) is
non-linear and we have an example of a dynamically non-linear theory
with a linear space of histories. In our terminology, this is an
example of a {\it kinematically linear} but {\it dynamically
non-linear} theory. Theories of non-abelian gauge fields, on the other
hand, are examples of theories in which non-linearities are present
already at the kinematical level.

In quantum field theory, the interest lies not in particular histories
satisfying (\ref{2.4}) but in summing over all histories, i.e.  in
defining measures on $\sh$ that correspond to the heuristic expression
\cite{GJ,VR,DI}
\be
d\mu(\phi) = ``{1 \over Z} e^{-S(\phi)} \prod_{x \in \R^{d+1}}
d \phi(x) "   \ .
\label{2.5}
\ee
We will begin in section 2.1 with a brief review of how measures are
constructed on infinite dimensional linear spaces. In section 2.2,
these techniques will be applied to two illustrative examples in
constructive quantum field theory; the massive free scalar field in
$d+1$ dimensions and the $\lambda\phi^4$-model in 2 dimensions.

\subsection{Integration on $\sh$}

We will first present an ``algebraic'' approach to integration and
then summarize the situation from a measure-theoretic viewpoint. In
the algebraic approach, the main idea is to reduce the problem of
integration over infinite dimensional spaces to a series of
integrations over finite dimensional spaces by judiciously choosing
the functions one wants to integrate.

In a linear space like $\sh $ the simplest functions that one can
introduce are the linear ones. Let $\S$ denote the space of all test
(or smearing) functions on $\R^{d+1}$, i.e.  the space of all
infinitely differentiable functions which fall off sufficiently
rapidly at infinity. $\S$ is called Schwarz space. Its elements $e \in
\S$ can be used to {\it probe} the structure of scalar fields
$\phi \in \sh$ through linear functions $F_e$ on $\sh$ defined by
\be
F_e(\phi) = \int_{\R^{d+1}}\ \phi(x) e(x)\ d^{d+1} x  \ ;
\label{2.6}
\ee
the test field $e$ probes the structure of $\phi$ because it captures
part of the information contained in $\phi$, namely, the ``component
of $\phi$ along $e$''. We can probe the behaviour of $\phi$ in the
neighbourhood of a point in $M$ by choosing test fields $e_n$
supported in that neighbourhood.

To begin with, we want to define integrals of such simple functions.
Let $e_1, ... , e_n$ denote arbitrarily chosen but fixed linearly
independent probes.  Consider the projection $p_{e_1, ...,e_n}$ they
define
\ba
p_{e_1, ... , e_n} \ : \ \sh \ & \rightarrow & \ \R^n  \nonumber \\
\phi   \ & \mapsto & \ \left(F_{e_1}(\phi), ... , F_{e_n}(\phi) \right)
\label{2.9} \ .
\ea
Next, consider functions $f$ on $\sh$ that depends on $\phi$ only
through their ``n-components'' $F_{e_1}(\phi), ... , F_{e_n}(\phi)$,
i.e. functions of the type
\be
\label{2.10}
f(\phi) = \widetilde f \left(F_{e_1}(\phi), ... , F_{e_n}(\phi) \right)
\ee
or equivalently
\be
f = p_{e_1, ... , e_n}^* \widetilde f   \ ,
\ee
where $\widetilde f$ is a (well-behaved) function on $\R^n$.  The
function $f$ is said to be {\it cylindrical} with respect to the
finite dimensional subspace $V_{e_1, ... , e_n}$ (of $\S$) spanned by
the probes $\left\{e_1, ... , e_n\right\}$. These are the functions
we first want to integrate.

This task is easy because the cylindrical functions are ``fake''
infinite dimensional functions: although defined on $\sh$, their
``true'' dependence is only on a finite number of variables.  Fix a
(normalized, Borel) measure $d\mu_{e_1, ... , e_n}$ on $\R^n$ and
simply define the integral of $f$ over $\sh$ to be the integral of
$\tilde{f}$ over $\R^n$ with respect to $d\mu_{e_1, ...,e_n}$:
\be
\label{2.11}
\int_{\sh} f(\phi) d\mu(\phi) := \int_{\R^n} \widetilde
f(\eta_1, ... , \eta_n)\ d\mu_{e_1, ... , e_n}(\eta_1, ... , \eta_n)
\  .
\ee
Next, in order to be able to integrate functions cylindrical with
respect to {\it any} finite dimensional subspace of the probe space,
we select, for every collection $\left\{e_1, ... , e_n\right\}$ of
linearly independent probes, and every $n \in \N$, a measure
$d\mu_{e_1, ... , e_n}$ on $\R^n$ and define the integral of that
cylindrical function over $\sh$ by (\ref{2.11}). This is the key
technique in the algebraic approach.

The procedure seems trivial at first sight. There is, however, a
catch. Since a function which is cylindrical with respect to a linear
subspace $V$ of $\sh$ is necessarily cylindrical with respect to a
linear subspace $V'$ if $V \subseteq V'$, the representation
(\ref{2.10}) is {\it not} unique. (Indeed, even for fixed $V$, the
explicit form of (\ref{2.10}) depends on the basis we use). For the
left side of (\ref{2.11}) to be well-defined, therefore, the finite
dimensional measures $\left(d\mu_{e_1, ... , e_n}\right)$ must satisfy
non-trivial consistency conditions; these serve precisely to ensure
that the integral of $f(\phi)$ over $\sh$ is independent of the choice
of a particular representation of $f$ as a cylindrical function. When
these consistency conditions are satisfied, $\sh$ is said to be
equipped with a {\it cylindrical measure}.

Let us consider a simple but representative example of consistency
conditions. Consider $e, \hat e \ \in \ \S$ and let $V_1, \ V_2$
be the one and two-dimensional spaces spanned by $e$ and
$e, \ \hat e$ respectively and $f$ be the  function on $\sh$,
cylindrical with respect to $V_1$, given by
\be
f(\phi) = \widetilde f_1 \left(F_e(\phi) \right)
:= \exp\ [i \l \int_{\R^{d+1}}\ e(x) \phi(x)\ d^{d+1}x]  \ .
\label{2.14}
\ee
This function is clearly cylindrical also with respect to $V_2$,
i.e. it is a function of $F_e $ and $F_{\hat e}$ that just happens
not to depend on $F_{\hat e}$:
\be
\label{2.15}
f(\phi) = \widetilde f_2 \left(F_e(\phi), F_{\hat e}(\phi) \right)
:= \exp\ [i \l \int_{\R^{d+1}}\ e(x) \phi(x)\ d^{d+1}x]  \ .
\ee
To obtain a cylindrical measure, therefore, $f, \ d\mu_e$,and
$d\mu_{e, \hat e}$ must satisfy:
\be
\int_\sh\ f(\phi)\ d\mu(\phi) = \int_\R e^{i \l \eta} d\mu_e(\eta)
= \int_{\R^2} e^{i \l \eta_1} d\mu_{e,\hat e}(\eta_1, \eta_2) \ .
\label{2.16}
\ee
It is easy to see that this equality holds for our choice of $f$ {\it
and} for any other integrable function, cylindrical with respect
to $V_1$, if and only if the measures satisfy the following
consistency condition
\be
\label{2.17}
d\mu_e(\eta) = \int_\R d\mu_{e,\hat e}(\eta, \hat \eta)   \ .
\ee
A natural solution to the consistency conditions is obtained by
choosing all the $d\mu_{e_1, ... , e_n}$ to be normalized Gaussian
measures. Then, the resulting $d\mu$ on $\sh$ is called a {\it
Gaussian cylindrical measure}.

Associated with every cylindrical measure (not necessarily Gaussian)
on $\sh$, there is a function $\chi$ on the Schwarz space ${\S}$ of
probes, called the Fourier transform of the measure by analysts and
the generating function (with imaginary current) by physicists:
\be
\label{2.18}
\chi(e) := \int_\sh\  \exp\ [i \int_{\R^{d+1}} e(x) \phi(x) d^{d+1}x]
\ d\mu(\phi) = \int_\R e^{i\eta} d\mu_e(\eta)   \ .
\ee
To see that $\chi$ is the generating functional used in the physics
literature, let us substitute the heuristic expression (\ref{2.5}) of
$d\mu$ in (\ref{2.18}) to obtain:
\ba
\label{2.19}
\chi(e) & = &`` {1\over Z} \int_\sh\
\exp[i\int_{\R^{d+1}} e(x) \phi(x)  d^{d+1}x ]\ \times \\
& & \exp[-\left(
\int_{\R^{d+1}}  {1 \over 2} \partial_\mu \phi(x) \partial^\mu \phi(x) +
V(\phi(x))
\right) d^{d+1}x ]
\prod_{x \in
\R^{d+1}} d\phi(x)" \nonumber\; .
\ea
 From the properties of these heuristic generating functionals
discussed in the physics literature, one would expect $\chi$ to
contain the complete information about $d\mu$. This is indeed the
case.

In fact, such generating functionals can serve as a powerful tool to
{\it define} non-Gaussian measures $d\mu$. This is ensured by a key
result in the subject, the {\it Bochner theorem} \cite{GV,Y}, which
has the following consequence :
\\
{\it Let $\chi(e)$ be a function on the Schwarz space
$\S$ satisfying the following three conditions} :
\ba \label{2.20}
&(i)& \chi(0) = 1 \nonumber \\
&(ii)& \chi \ \hbox{{\rm is continuous in every finite dimensional
subspace of}} \S \ \nonumber \\
&(iii)& \hbox{For every $e_1, ... , e_N \in \S$ and $c_1, ... , c_N
\in \C$ we have}
\nonumber \\
&& \sum_{i,j=1}^N \overline {c_i} c_j \chi(-e_i + e_j) \ge 0 \ .
\ea
{\it Then, there exists a unique cylindrical measure $d \mu$ on $\sh$ such
that $\chi$ is its generating functional}.
\smallskip \noindent
Thus, functions $\chi$ satisfying (\ref{2.20}) are generating
functionals of cylindrical measures on $\sh$ and every generating
functional is of this form.  This concludes the ``algebraic'' part of
our discussion.

We now turn to the measure-theoretic part and ask if the above
procedure for integrating cylindrical functions actually defines a
genuine measure --i.e., a $\sigma$-additive set function (see below)--
on $\sh$ or a related space. This issue is important because a proper
measure theoretic understanding would enable us to integrate functions
on $\sh$ that are genuinely infinite dimensional, i.e., depend on
infinitely many probes. Indeed, the classical action is invariably
such a function.  It turns out that one {\it can} define genuine
measures, but to do so, we have to extend the space $\sh$. For later
convenience, we will proceed in two steps. First, we will present a
``maximal'' extension and arrive at a space $\shb$ which serves as
``the universal home'' for all cylindrical measures on $\sh$. In
practice, however, the space $\shb$ is too large in that the measures
of interest to constructive quantum field theory are generally
supported on a significantly smaller subspace $\S'$ of $\shb$ (which
are still larger than $\sh$). Then, in the second step, we will
discuss this ``actual home,'' $\S'$, for physically interesting
measures.

The universal home, $\shb$, is simply the algebraic dual of $\sh$: the
space of {\em all} linear functionals on the probe space $\S$. This
space is ``very large'' because we have not required the maps to be
continuous in any topology on $\S$. It is easy to check that $\S$ also
serves as the space of probes for $\shb$ and that, given a consistent
family of measures $d\mu_{e_1, ...,e_n}$ that defines a cylindrical
measure $d\mu$ on $\sh$, it also defines a cylindrical measure, say
$d\bar\mu$, on $\shb$. ($\shb$ is the ``largest'' space for which this
result holds.) Now, for a cylindrical measure to define a genuine,
infinite dimensional measure, it has to be extendible in the following
sense.  For a cylindrical measure, the measurable sets are all
cylindrical, i.e., inverse images, under projections $p_{e_1, ...
,e_n}$ of (\ref{2.9}), of measurable sets in $(\R^n, d\mu_{e_1, ...
,e_m})$.  The measure of a cylindrical set is just the measure of its
image under the projection.  Now, for $d\bar\mu$ to be a genuine
measure, it has to be $\sigma$-additive, i.e., the measure of a {\it
countable} union of non-intersecting measurable sets has to equal the
sum of their measures. Unfortunately, although the union of a finite
number of cylindrical sets is again cylindrical, in general the same
is {\it not} true for a countable number. Thus, the question is
whether we can add countable unions to our list of measurable sets and
{\it extend} $d\bar\mu$ consistently. It turns out that {\it every}
cylindrical measure on $\shb$ can be so extended. (This is in general
{\it not} true for $\sh$; even the Gaussian cylindrical measures on
$\sh$ may not be extendable to $\s$-additive measures thereon. In
particular this happens for physically interesting measures.) This is
why $\shb$ can be regarded as the ``universal home'' for cylindrical
measures.

Unfortunately, the space $\ov \sh$ is typically too big for quantum
scalar field theories: The {\it actual} home of a given measure $d
\bar \mu$ is generally a smaller space, say $\wh \sh$, of
better behaved generalized histories, in the sense that $\bar\mu (\ov
\sh - \wh \sh) = 0$. (More precisely, every measurable set $U$
such that $U \subset \ov \sh - \wh \sh$ has zero measure $\bar\mu(U) =
0$.) Since $\wh \sh$ has a richer structure, it is most natural
(and, in practice, essential) to ignore $\shb$ and work directly with
$\wh\sh$.

The key result which helps one determine the actual home of a measure
is the Bochner-Minlos theorem \cite{Y}. The version of this theorem
which is most useful in scalar field theories can be stated as follows
:\\ {\it $\s$-additive measures on the space $\S'$ of {\it continuous}
linear functionals on the probe space $\S$ (equipped with its natural
nuclear topology $\tau_{(n)}$) are in one to one correspondence with
generating functions $\chi$ on $\S$, satisfying the conditions (i),
(iii) of (\ref{2.20}) and
\be
(ii') \chi \mbox{ is continuous with respect to } \tau_{(n)}.
\ee
}
\smallskip\noindent
(The space $\S'$ is of course the space of tempered distributions.)
Since the topology on $\S$ used in $(ii')$ is weaker than that in
$(ii)$, the Bochner-Minlos theorem expresses the general trend that
the {\it weaker} the topology with respect to which the generating
function $\chi$ is continuous, the {\it smaller} is the support of the
resulting measure $d\bar\mu$.

This concludes our discussion of mathematical preliminaries.  Let us
summarize. One can define cylindrical measures on $\sh$ which enable
one to integrate cylindrical functions. However, to integrate
``genuinely infinite dimensional'' functions, one needs a genuine
measure. The algebraic dual $\shb$ of the space $\S$ of probes is the
univeral home for such measures in the sense that every cylindrical
measure on $\sh$ can be extended to a genuine measure on $\shb$. In
practice, however, physically interesting measures have a much smaller
support $\wh\sh$ which, however, is larger than $\sh$. (Indeed,
typically $\sh$ has {\it zero} measure.) The Bochner-Milnos theorem
provides a natural avenue to constructing such measures.

\subsection{Scalar field theories}

For a measure $d \bar\mu$ to correspond to a physically interesting
quantum scalar field theory, it has to satisfy (some version) of the
Osterwalder-Schrader axioms \cite{GJ,VR}. These axioms guarantee
that from the measure it is possible to construct the physical Hilbert
space with a well defined Hamiltonian and Green functions with the
appropriate properties. They are based on the assumption that the
actual home for measures that correspond to quantum scalar field
theories is the space $\S'$ of tempered distributions. Thus, the
appropriate histories for quantum field theories are distributional.
In fact, typically, the measure is {\it concentrated} on genuine
distributions in the sense that $\bar{\mu}(\S' - \sh) =1$. This is
the origin of ultra-violet divergences: while the measure is
concentrated on distributions, the action (\ref{2.3}) is ill-defined
if the histories are distributional.

We will denote the measures on $\S'$ by $d\hat\mu$.  The
Osterwalder-Schrader axioms restrict the class of possible measures.
The most important of these axioms are: Euclidean invariance and
reflection positivity. The first requires that the measure $d\hat\mu$
be invariant under the action of the Euclidean group on $\R^{d+1}$.
Reflection positivity is the axiom that allows to construct a physical
Hilbert space with a non-negative self-adjoint Hamiltonian acting on
it.  Let $\theta$ denote the time reflection, i.e. reflection with
respect to the hyperplane
\be
(x^0 = 0, x^1, ... , x^d) \ .
\ee
Consider the subspace ${\cal R}^+$ of $L^2(\S', d \wh \mu)$ of
(cylindrical) functions of the form
\be
\label{2.36}
f(\hat\phi) = \sum_{j=1}^{N} c_j e^{i \hat\phi (e_j^+)}  \ ,
\ee
where $c_j \in \C$ and $e_j^+$ are arbitrary probes with support in
the $x^0 > 0$ half-space. Then, the reflection positivity of the
measure is the condition that
\be
< \theta f , f >_{L^2} = \int_{\S'} [(\theta f)(\hat\phi)]^\star
\; f(\hat\phi)\; d \hat\mu (\hat \phi) \ \ge 0 \ . \label{2.37}
\ee
(If reflection invariance is satisfied for one choice of Euclidean
coordinates, by Euclidean invariance of the measure, it is satisfied
for any other choice.)

The Hilbert space and the Hamiltonian can then be constructed as
follows. (\ref{2.37}) provides a degenerate inner product $(.,.)$ on
${\cal R}^+$, given by
\be
\label{2.38}
(f_1, f_2) = <\theta f_1, f_2>_{L^2}    \ .
\ee
Denoting by $\cal N$ the subspace of $(.,.)$-null vectors on ${\cal
R}^+$ we obtain a Hilbert space $\cal H$ by taking the quotient
${\cal R}^+/\cal N$ and completing it with respect to $(.,.)$:
\be
\label{2.39}
{\cal H} =  \overline{{\cal R}^+/\cal N}    \ .
\ee
The Euclidean invariance provides an unitary operator $\wh T_t \ , \ t
> 0 $, of time translations on $L^2(\S', d \hat\mu)$. It in turn gives
rise to the self-adjoint contraction operator on ${\cal H}$:
\be
\label{2.40} e^{-Ht} \ , \ t > 0 \ , \ee where $H$ is the Hamiltonian.

Finally, we can formulate the two axioms in terms of the generating
functional.  Let $ E \ : \ \R^{d+1} \ \rightarrow \ \R^{d+1}$ denote
an Euclidean transformation. Then the condition that the measure be
Euclidean invariant is equivalent to demanding:
\be
\hat{\chi}(e \circ E) = \hat{\chi} (e)   \ , \ \hbox{{\rm for every}}
\ E   \ .
\label{2.34}
\ee
Next, let us consider reflection positivity. From the definition
of the generating functional $\hat{\chi}$ (\ref{2.18}) (here with
${\cal SH}\equiv\S'$) it follows that the
condition of reflection positivity is equivalent to
\be
\sum_{i,j =1}^N \ov c_i c_j \hat{\chi}(e_j^+ - \theta e_i^+)  \ \ge \ 0
\label{2.42}
\ee
for all $N \in \N, \ c_1, ... , c_N \ \in \ \C$ and for all $e_1^+
, ... , e_N^+ \ \in
\ {\cal R}^+$.

To summarize, constructive quantum field theory provides an elegant
and compact characterization of quantum field theory as a measure
$d\hat\mu$ on $\S'$ or as a (generating) function $\hat{\chi}$ on the space
$\S$ of test functions, satisfying certain conditions.  So all the
work can be focussed on finding (or at least proving the existence of)
appropriate $d\hat\mu$ or $\hat{\chi}$.

To conclude this section, we will provide two examples of such
measures.

The first example is that of a free, massive scalar field on
$\R^{d+1}$. Note first, that, it follows from the discussion in
section 2.1 that a measure $d\hat\mu$ is Gaussian if and only if its
generating functional is Gaussian, i.e. if and only if
\be
\label{2.21}
\hat{\chi} (e) = \exp[-{1 \over 2}(e, \widehat C e)] \ ,
\ee
where $\widehat C$ is a positive definite, linear operator defined
everywhere on $\S$. $\widehat C$ is called the {\it covariance} of the
resulting Gaussian measure $d\hat\mu$.  The free massive quantum
scalar field theory corresponds to the Gaussian measure $d\hat\mu_m$
with covariance
\be
\label{2.24}
\left( \widehat C_m e \right)(x) = \int_{\R^{d+1}}
(- \Delta + m^2 )^{-1}(x,y)\ e(y)\ d^{d+1}y   \ ,
\ee
where the integral kernel is defined by:
$$
(- \Delta + m^2 )^{-1}(x,y) = {1 \over (2\pi)^{d+1}}
\int_{\R^{d+1}}\ d^{d+1}p \ {e^{ip(x-y)} \over p^2 + m^2}.
$$

Our next example is more interesting in that it includes interactions:
the $\lambda\phi^4$ model in 2-dimensions. The classical action is now
given by:
\be
\label{2.43}
S(\phi) = \int_{\R^2} \left( {1 \over 2} \partial_\mu \phi
\partial^\mu \phi + {m^2 \over 2} \phi^2 + \l \phi^4 \right) d^2x
\ee
Therefore, the heuristic expression for the quantum measure is
\ba
d \mu_{\l, m} &=& {1 \over Z_{\l, m}}\ e^{-\l \int_{\R^2} \hat\phi^4(x)
d^2x}\ e^{-S_m(\hat\phi)} \prod_{x \in \R^2} d \hat\phi(x)
\nonumber \\
&=& {1 \over \wt Z_{\l,m}} \left( e^{-\l \int_{\R^2} \hat\phi^4(x) d^2x}
\right)\ d\hat{\mu}_m(\hat\phi )  \ ,
\label{2.44}
\ea
where $S_m$ is the action of the free field with mass $m$ and
$d\hat{\mu}_m$ denotes the (rigorous) Gaussian measure discussed
above. The problem with this expression is that while $d\hat\mu$ is
concentrated on distributional connections, the factor in the exponent
is ill-defined for distributional $\hat\phi$.

One way of trying to make sense of (\ref{2.44}) is by substituting the
Gaussian measure $d\hat{\mu}_m$ by a regulated Gaussian measure with a
smaller support on which the integrand is well-defined. This can be
achieved, e.g., by replacing the covariance $C_m$ in (\ref{2.24}) with
the cutoff covariance $C_m^k$, given in the momentum space by
\cite{VR}
\be
\wt C_m^k(p) = \frac{1}{(2\pi)^2}{ e^{-k(p^2 + m^2)} \over p^2 + m^2},
\; k>0 \ .\label{2.45}
\ee
Then the resulting measure $d \mu^k_m$ ``lives" on the space ${\cal
L}$ of {\it infinitely differentiable} functions that grow at most
logarithmically at infinity (more precisely as $\sqrt{\ln(|x|)}$).
(This is an extremely useful property. Nonetheless, we cannot just use
this measure as the physical one because, among other reasons, it does
not satisfy reflection positivity.) However, the function
\be
\label{2.46}
\l \int_{\R^2} \phi^4(x) d^2x
\ee
is still infrared divergent (almost everywhere with respect to
$d\mu_m^k$). Therefore, we have to put an infrared cutoff by
restricting space-time to a box of volume $V$. Of course if we now
take the regulators away, $V \rightarrow \infty$ and $k \rightarrow
0$, then we return to the ill-defined expression (\ref{2.44}).
Therefore, the question is whether it is possible to change $\l
\int_{\R^2} \phi^4(x) d^2x$ by an interacting term with the same
leading order dependence on $\phi$ but such that the limit, when the
regulators $V \rightarrow \infty, \ k \rightarrow 0$, exists, is non
trivial and satisfies the Osterwalder-Schrader axioms.

In $2$-dimensions the answer is in the affirmative if we substitute
(\ref{2.46}) by
\cite{VR}
\be
\label{2.47}
\l \int_{\R^2} :\phi^4(x): d^2x    \ ,
\ee
where $:\phi^4(x):$ denotes normal ordering with respect to the
Gaussian measure $d \mu_m^k$
\be
\label{2.48}
 :\phi^4(x): \ = \ \phi^4(x) - 6 C_m^k(0) \phi^2(x) + 3 (C_m^k(0))^2   \ .
\ee
The expression for the regulated generating functional then reads
\be
\label{2.49}
\chi_{m,\l}^{k,V} := {1 \over \wt Z^{k,V}_{\l, m}}
\int_{\cal L}\ \exp[i \int_{V} e \phi d^2x]\
\exp[-\l \int_{V} : \phi^4(x): d^2x]\ d\mu^k_m(\phi )  \ ,
\ee
where ${\cal L}$ denotes the support of the measure (i.e., the space of
$C^\infty$ functions on $\R^2$ which grow at worst as $\sqrt{ln(|x|)}$).
Finally, it can be now shown that the limit
\be
\label{2.50}
\chi_{m,\l}(e) = \lim_{V \rightarrow \infty}
\lim_{k \rightarrow 0} \chi_{m,\l}^{k,V}(e)
\ee
exists, is non-Gaussian and has the appropriate properties. Thus,
while the extraordinarily difficult problem of rigorously constructing
a quantum field theory is formulated succinctly in this approach as
that of finding a suitable measure or generating functional, the
actual task of finding physically interesting measures is
correspondingly difficult.

Finally, note that the entire framework is ``soaked in'' kinematic
linearity. The fact that the space $\S$ of probes and and home $\S'$
of measures are linear was exploited repeatedly in various steps.

\section{Calculus on $\ag$}

We now turn to gauge theories which are kinematically non-linear.
Here, the classical space of histories is the infinite dimensional
space $\ag$ of smooth connections modulo gauge transformations on the
$d+1$ dimensional space-time $M$; ${\cal SH}=\ag$.  We will assume the
gauge group $G$ to be a compact Lie group and, in specific
calculations, take it to be $SU(2)$. In this section, we will
summarize how one develops calculus on $\ag$ by suitably extending the
arguments of section 2.1 \cite{AL1,AL2}.  (Analogous methods for $d+1
= 2$ were first used in the second reference of \cite{GKK}. For an
alternative approach, see \cite{As}).  In the section 4, we will apply
these ideas to a specific model along the lines of section 2.2

\subsection{Cylindrical measures}

The main idea (\cite{AI,AL1}) is to substitute the linear duality
between scalar field histories and test functions by the ``non-linear
duality'' between connections and loops in M. This duality is provided
by the parallel transport or holonomy map H, evaluated at a base point
$p\in M$:
\be \label{3.2}
H:\; (\alpha,A)\to H_\alpha(A)={\cal P}\exp(\oint_\alpha\ A.dl)\;
\in \; G\ , \ee
where $\alpha$ is a piecewise analytic loop in M and $A\in\cal A$.
(For definiteness we will assume that $A$ is explicitly expressed
using (one of) the lowest dimensional representation(s) of the Lie
algebra of G.)  In view of this duality, loops will now be used to
probe the structure of the space of connections. In the kinematically
linear theories, the probes and the histories were objects of the same
type; in the scalar field theories, for example, they were both
functions on $M$. In gauge theories, the roles are played by quite
different objects.

We will begin by specifying the precise structure of the space of
probes.  Fix a point $p\in M$ and consider only those loops $\alpha$
in $M$ which are based at $p$. It is natural to define the following
equivalence relation on the space of these loops:
\be \label{3.3}
\alpha'\sim \alpha \mbox{ iff }H_{\alpha'}(A)=H_\alpha(A),\;
\forall A\in\cal A\ ,
\ee
where $\cal A$ is the space of smooth connections on $M$. Denote the
equivalence class by $[\alpha]_p$ and call it a (based) holonomic loop
or a {\it hoop}. The set of all hoops, $\hgp$, in $M$ forms a group
with respect to the product
\be \label{3.4}
[\alpha_1]_p\cdot [\alpha_2]_p = [\alpha_1\circ_p\alpha_2]_p\ ,
\ee
where $\alpha_1\circ_p\alpha_2$ denotes the usual composition of loops
at the base point $p$. For gauge theories, the hoop group $\hgp$ will
serve as the space of probes.

Let us now turn to connections. For simplicity, we will first consider
the space $\agp$ where ${\cal G}_p$ is the subgroup of gauge
transformations which are equal to the identity $1_G\in G$ at $p$.
For each $\alpha\in\hgp$ the holonomy $H_\alpha$ defines a G-valued
function on $\agp$, i.e. a G-valued function on $\cal A$ which is
invariant under ${\cal G}_p$-gauge transformations. These functions
are sufficient to separate the points of $\agp$. That is, for every
$[A_1]_p\not=[A_2]_p$ there exists a $\alpha\in\hgp$ such that
$H_\alpha(A_1)\not=H_\alpha(A_2)$, where $[A]_p$ denotes the ${\cal
G}_p$ equivalence class of the connection $A$. In this sense, hoops
play the role of non-linear probes for histories $[A]_p\in
\agp$. Finally, note that each smooth connection $A\in\cal A$ defines
a homomorphism $H$ of groups,
\be \label{3.8} H_.(A):\;\hgp\to G;\;\ \alpha\to H_\alpha(A), \ee
which is smooth in an appropriate sense (\cite{B}).

As in section 2.2, we can now define cylindrical functions and
cylindrical measures on $\agp$ using the ``probe functions''
$F_\alpha$ defined by $F_\alpha([A]_p) := H_\alpha(A)$. Our first task
is to introduce the analogs of the projections ({\ref{2.9}). For this,
we need the notion of strongly independent hoops. Following
\cite{AL1}, we will say that hoops $[\beta_i]_p$ are strongly
independent if they have representative loops $\beta_i$ such that each
contains an open segment that is traced exactly once and which
intersects other representatives at most at a finite number of
points. The notion of strong independence turns out to be the
appropriate non-linear analog of the linear independence of probes
used in section 2.1. In particular, we have the following result.
Given $n$ strongly independent hoops, $[\beta_i]_p, i=1,... ,n$, there
exists a set of projections, $p_{\beta_1,...,\beta_n}: \agp\to
G^n$, given by:
\be \label{3.10}
p_{\beta_1,..,\beta_n}
([A]_p) = (H_{\beta_1}(A),..,H_{\beta_n}(A))
\ee
which are surjective \cite{AL1}.

Now, a function f on $\agp$ is called {\it cylindrical} with respect
to the subgroup of $\hgp$ generated by $\beta_1,..,\beta_n$ if it is the
pull-back by $p_{\beta_1,..,
\beta_n}$ of a (well-behaved) complex valued function $\tilde{f}$ on
$G^n$
\be \label{3.11}
f([A]_p)=\tilde{f}(H_{\beta_1}(A),..,H_{\beta_n}(A)).
\ee
Consider now a family \{$d\mu_{\beta_1,..,\beta_n}$\} of (positive,
normalized) measures on $G^n$, one for each set $\{\beta_1,..,
\beta_n\},\;n\in \N$ of strongly independent hoops. As in section 2.1,
this family of measures on finite dimensional spaces allows us to
define in a unique way a {\it cylindrical measure} $d\mu$ on the
infinite dimensional space $\agp$, provided that appropriate
consistency conditions are satisfied. These conditions are again a
consequence of the fact that the representation (\ref{3.11}) of $f$ as
a cylindrical function is not unique. If the consistency conditions
are satisfied, then the family $(d\mu_{\beta_1,..,\beta_n})$ defines a
unique cylindrical measure $d\mu$ on $\agp$ through
\be \label{3.12}
\int_{\agp} d\mu([A]_p) f([A]_p):=\int_{G^n}
d\mu_{\beta_1,..,\beta_n}(g_1,..,g_n)
\tilde{f}(g_1,..,g_n)
\ee
where $f$ is the pull-back of $\tilde{f}$ as in (\ref{3.11}).

In the linear case, Gaussian measures on $\R^n$ provided a natural way
to meet the consistency conditions. In the present case, one can use
the Haar measure on $G^n$. More precisely, the consistency conditions
are satisfied if one chooses
\cite{AL1}
\be \label{3.13}
d\mu^0_{\beta_1,..,\beta_n}(g_1,..,g_n)=d\mu_H(g_1)..d\mu_H(g_n)\ ,
\ee
where $d\mu_H$ is the normalized Haar measure on $G$.  This family
$(d\mu^0_{\beta_1,..,\beta_n})$ leads to a measure cylindrical
$d\mu^0$ on $\agp$ which is in fact invariant under the action of
diffeomorphisms
\footnote{Additional solutions of the consistency conditions
leading to diffeomorphism invariant measures on $\agp$ were found by
Baez in \cite{B2}. The resulting measures are sensitive to certain
kinds of self-intersections of loops. Such measures are relevant for
quantum gravity, formulated as a dynamical theory of connections
\cite{A1},\cite{A2}.} of $M$.

\subsection{``Universal home'' and ``actual home'' for measures}

As in the linear case, not every cylindrical measure on $\agp$ is a
genuine, infinite dimensional measure. It turns out, however, that
they can be extended to genuine measures on a certain completion,
$\agpb$.  This comes about as follows.  Since $\agp$ is in one to one
correspondence with the set of smooth homomorphisms from $\hgp$ to
$G$ \cite{B}, a natural analog of the algebraic dual $\shb$ of the space
$\S$ of linear probes is now
\be \label{3.14}
\agpb:=Hom(\hgp,G),
\ee
i.e., the set of {\em all } homomorphisms (without any continuity
condition) from the hoop group to the gauge group \cite{AL1}. Elements
of $\agpb$ will be denoted by $\bar{A}$ and called {\it generalized
connections}. Just as $\S$ continues to serve as the space of probes
for $\shb$ in the linear case, the group $\hgp$ continues to provide
non-linear probes for the space $\agpb$ of generalized connections.
Furthermore, as in the linear case, every consistent family
$(d\mu_{\beta_1,..,\beta_n})$ of measures on $G^n$ that defines a
cylindrical measure $d\mu$ on $\agp$ through (\ref{3.12}) also defines
a measure $d\bar{\mu}$ on $\agpb$ by
\be \label{3.17}
\int_{\agpb} \bar{f}(\bar{A}) d\bar{\mu}(\bar{A})
=\int_{G^n} \tilde{f}(g_1,..,g_n) d\mu_{\beta_1,..,\beta_n}(g_1,..,g_n)
\ee
where $\bar{f}$ is an arbitrary cylindrical function on $\agpb$ with
respect to the subgroup of $\hgp$ generated by $\beta_1,..,\beta_n$;
$\bar{f}(\bar{A})=\tilde{f}(\bar{A}(\beta_1),..,\bar{A} (\beta_n))$.
Finally, every measure $d\bar{\mu}$ on $\agpb$ defined as in
(\ref{3.17}) can be extended to a $\sigma$-additive measure on $\agpb$
and is thus a genuine, infinite dimensional measure \cite{Y,
AL1, MM}.

 From a physical point of view, however, we should still factor out by
the gauge freedom {\em at} the base point $p$.  A generic gauge
transformation $g(.)\in \cal G$ acts on $\agpb$ simply by
conjugation
\be \label{3.19}
(g\circ\bar{A})(\beta)=g(p)\cdot\bar{A}(\beta)\cdot g^{-1}(p)
\ee
The physically relevant space $\agb$ is therefore the quotient of
$\agpb$ by this action.  The classical space of histories $\ag$ is
naturally embedded in $\agb$ through
\be \label{3.21} [A]\to\{g H_.(A) g^{-1},\; g\in G\} \ee
where $[A]$ denotes the gauge-equivalence class of the connection
$A$. By integrating gauge-invariant functions on $\agpb$ with the
help of $d\bar{\mu}$ we obtain a measure on $\agb$ that we will denote
also by $d\bar{\mu}$. This measure is of course just the push-forward
of the measure on $\agpb$ under the canonical projection $\pi\; :\;
\agpb\to\agb $. Again, these measures $d\bar{\mu}$ on $\agb$,
associated with consistent families $(d\mu_{\beta_1,..,\beta_n})$, are
always extendible to $\sigma-$additive measures. Thus, for gauge
theories, $\agb$ serves as the universal home for measures (for which
the traces of holonomies are measurable functions) in the same sense
that $\shb$ is the universal home for measures in the linear case.

It is therefore natural to ask if there is a ``non-linear'' analog
of the Bochner theorem. The answer is in the affirmative. In fact the
arguments are now simpler because both $\agpb\mbox{ and }\agb$ are
compact Hausdorff spaces (with respect to the natural Gel'fand
topologies). To see this explicitly, let us now restrict ourselves to
the gauge group $G=SU(2)$. In this case, the Mandelstam identities
imply that the vector space generated by traces of holonomies
${T}_\alpha$ is closed under multiplication. This in turn implies that
the entire information about the measure is contained in the image
of the ``loop transform:''
\be \label{3.25}
\chi(\beta)=\int_{\agb}\ \bar{T}_\beta([\bar{A}])\
d\bar{\mu}([\bar{A}])\ ,
\ee
where $\bar{T}_\alpha [\bar{A}]=\frac{1}{2} {\rm tr} \bar{A}[\alpha]$ is the
natural extension to $\agb$ of the ``trace of the holonomy (or Wilson
loop) function''
on $\ag$. Note that (\ref{3.25}) is the natural non-linear analog of
the Fourier transform (\ref{2.18}). The generating function $\chi$
is again a function on the space of probes which now happens to be
the hoop group $\cal HG$ rather than the Schwarz space $\S$. From the
normalization and the positivity of $d\bar{\mu}$ it is easy to see
that $\chi(\alpha)$ satisfies
\ba \label{3.26}
(i) & & \chi (p)=1\nonumber \\
(ii) & & \sum_{{i,j}=1}^N \bar{c}_i c_j[\chi (\beta_i\circ\beta_j)
+\chi (\beta_i\circ\beta_j^{-1})]\ge 0,\ \forall \beta_i \in \hg\ ,
\ea
where $p$ denotes the trivial (i.e., identity) hoop, $c_i\in \C$, are
arbitrary complex numbers and $ N\in \N$ is an arbitrary integer.
Finally, the Riesz-Markov theorem \cite{ReeSi} implies that every
generating functional $\chi$ satisfying $\sum_i c_i \chi(\beta_i)
=0\mbox{ whenever } \sum_i c_i T_{\beta_i}=0$ is the loop transform of
a measure $d\bar{\mu}$ so that there is a one to one correspondence
between positive, normalized (regular, Borel) measures on $\agb$ and
generating functionals on $\hgp$ \cite{AI}. This result is analogous
to the Bochner theorem in the linear case.

As in section 2.2, a given measure $\bar{\mu}_0$ on $\agb$ may be
supported on a smaller space of better behaved generalized connections
(support or actual home for $d\bar{\mu}_0$). Indeed, just as $\shb$ is
``too large'' for scalar field theories, we expect that $\agb$ is
``too large'' for Yang-Mills theories. To see this, note that in
(\ref{3.26}), no continuity condition was imposed on the generating
functional $\chi_{\bar{\mu}_0}$ (or, equivalently, the generating
functionals are assumed to be continuous only with the discrete
topology on $\hgp$).  Now, in Yang-Mills theories (with $d+1>2$), a
background space-time metric {\it is} available and it can be
used to introduce suitable topologies on $\hgp$ which would be much
weaker than the discrete topology.  It would be appropriate to require
that the Yang-Mills generating functional $\chi$ be continuous with
respect to one of these topologies. Now, from our experience in the
linear case, it seems reasonable to assume that the following pattern
will emerge: the weaker the topology with respect to which $\chi$ is
continuous, the smaller will be the support of $d\bar\mu$. Thus, we
expect that, in Yang-Mills theories, physically appropriate continuity
conditions will have to be imposed and these will restrict the support
of the measure considerably. What is missing is the non-linear analog
of the Bochner-Minlos theorem which can naturally suggest what the
domains of the physically interesting measures should be.

The situation is very different in diffeomorphism invariant theories
of connections such as general relativity. For these theories,
diffeomorphism invariant measures such as $d\bar{\mu}^0$ are expected
to play an important role. There are indications that the generating
functionals of such measures will be continuous only in topologies
which are much stronger than the ones tied to background metrics on
$M$.  Finally, since $d\bar{\mu}^0$ is induced by the Haar measure on
the gauge group, it is very closely related to the measure used in the
lattice gauge theories. Therefore, it is possible that even in
Yang-Mills theories, it may serve as a ``fiducial'' measure
--analogous to $\mu^k_m$ of the scalar field theory-- in the actual
construction of the physical measure. We will see that this is indeed
the case in $d+1=2$.

\section{Example : Quantum Yang Mills Theory in two dimensions}

In this section we will show how the mathematical techniques
introduced in section 3 can be used to construct Yang-Mills theory in
2 dimensions in the cases when the underlying space-time $M$ is
topologically $\R^2$ and when it is topologically $S^1\times\R$.  In
particular, we will be able to show equivalence
rigorous
\footnote{Similar results were obtained via the heuristic Fadeev-Popov
approach in \cite{DF}.}
between the Euclidean and the Hamiltonian theories; to our knowledge
this was not previously demonstrated. The analysis of this case will
also suggest an extension of the Osterwalder-Schrader axioms for gauge
theories.  Due to space limitations, however, we will present here
only the main ideas; for details see \cite{ALMMT}.

\subsection{Derivation of the continuum measure}

We wish to construct the Euclidean quantum field theory along the
lines of section 2.2.  As indicated in section 3.2, the analog of the
generating functional $\chi(e)$ on $\S$ is the functional
$\chi(\alpha)$ on the hoop group $\hgp$. In the present case, the
heuristic expression of $\chi(\alpha)$ is:
\be \label{4.1}
``\chi(\alpha) = \int_{\ag}\ T_{\alpha}([A])\ e^{-S_{\rm YM}([A])}\
\prod_{x\in M}d[A](x)'' \ ,
\ee
where $[A]$ is the gauge equivalence class to which the connection $A$
belongs, $T_\alpha([A])$ is the trace of the holonomy of $A$ around the
closed loop $\alpha$ and $S_{YM}$ is the Yang-Mills action. To obtain
the rigorous analog of $(\ref{4.1})$, we proceed, as in section 2.2,
in the following steps: i) regulate the action using suitable
cut-offs; ii) replace $\ag$ by a suitable completion thereof; iii)
introduce a measure on this completion with respect to which the
regulated action is measurable; iv) carry out the integration; and, v)
take the appropriate limits to remove the cut-offs.

To carry out the first step, we will use a lattice regularization.
Introduce a finite, square lattice of total length $L_x
\mbox{ and } L_\tau$ in $x$ and $\tau$ direction respectively, and lattice
spacing $a$, using the Euclidean metric on $M$. Thus we have imposed
both infrared and ultraviolet regulators. There are $(N_x+1)(N_\tau
+1)$ vertices on that finite lattice in the plane (and $(N_\tau+1)N_x$
in the cylinder) where $N_x a:=L_x,\; N_\tau a:=L_\tau$. Denote the
holonomy associated with a plaquette $\Box$ by $p_\Box$. There are no
boundary conditions for the plane while on the cylinder we identify
\cite{C} the vertical link variables $l$ associated with the open
paths starting at the vertices $(0,\tau)$ and $(a \ N_x,\tau)$. The
regularized action is then given by $S_{\rm reg} = \beta S_{W}$ where
$\beta = 1/(g_0 a^2)$, with $g_0$, the bare coupling constant, and
$S_{W}$ is the Wilson action given by:
\be \label{4.3}
S_{W}:=\sum_\Box [1-\frac{1}{N} \mbox{Re tr}(p_\Box))]\ ,
\ee
where $\mbox{Re tr}$ stands for ``real part of the trace.''  Finally,
we note certain consequences of this construction. Choose a vertex and
call it the base point $p$. One can show that in two dimensions the
based plaquettes form a complete set of loops which are independent in
the sense that one can separately assign to each holonomy associated
with a plaquette an arbitrary group element. This means that they can,
in particular, be used as independent integration variables on the
lattice. They can also be used to define projection maps of
(\ref{3.10}).

The next step is to find the appropriate extension of $\ag$ and a
measure thereon. For this, we note that since $\agb$ of section 3.2 is
the space of all homomorphisms from $\hgp$ to $G$, for any given
lattice, the Wilson action (\ref{4.3}) can be regarded as a bounded
cylindrical function
$\bar{S}_{W}$ on $\agb$. Therefore, this function is integrable
with respect to the measure $d\bar{\mu}^0$. Since the
push-forward of $d\bar\mu^0$ under the projections $p_{\beta_1,...
,\beta_n}$ of (\ref{3.10}) is just the Haar measure on $G^n$, using for
$\beta_i$ the plaquettes, we have:
\ba \label{4.2}
\chi^{a,L_x,L_\tau}\ (\alpha) &:=&
\frac{1}{Z} \int_{\agb} \ \bar{T}_\alpha (\bar{A}) \  e^{-\beta
\bar{S}_W}[\bar{A}] \ d\bar{\mu}^0 \\
&=&\frac{1}{NZ(a,L_x,L_\tau)}
\prod_l\ \int_G
\ e^{-\beta S_{W}} \ \mbox{tr}(\prod_{l\in\alpha} H_l) \
d\mu_H(H_l) \nonumber
\ea
for any loop $\alpha$ contained in the lattice. Here $d\mu_H$ is the
Haar measure on $G$ and the partition function $Z$ is defined through
$\chi(p) = 1$ where, as before, $p$ denotes the trivial hoop.  Note that
(\ref{4.2}) is precisely the Wilson integral for computing the vacuum
expectation value of the trace of the holonomy. Thus, the space $\agb$
and the measure $d\bar{\mu}^0$ are tailored to the calculations one
normally performs in lattice gauge theory.

We have thus carried out the first four steps outlined above. The last
step, taking the continuum limit, is of course the most difficult one.
It has been carried out for the general case $G= SU(N)$ \cite{ALMMT}.
For simplicity, however, in what follows, we will restrict ourselves
to the Abelian case $G=U(1)$.

Consider the pattern of areas that the loop $\alpha$ creates on $M$
and select simple loops $\beta_i$ that enclose these areas. It can be
shown that $\alpha$ can be written as the composition of the $\beta_i$
and the completely horizontal loop $c$ at ''future time infinity".
Let $k(\beta_i)\mbox{ and }k$ be the effective winding numbers of the
simple loops $\beta_i,\; i=1,..,,n$ and of the homotopically
non-trivial loop $c$ respectively, in the simple loop decomposition of
$\alpha$ (that is, the signed number of times that these loops appear
in $\alpha$). Define $|\beta_i|$ to be the number of plaquettes
enclosed by $\beta_i$.  Then, if we set
\be
K_n(\beta):=\int_G \  [\exp-\beta(1-\mbox{Re}(g))]\
\ g^n \ d\mu_H(g),
\ee
the generating functional (on the plane or the cylinder) becomes
\be \label{4.4}
\chi(\alpha)=\prod_{i=1}^n\ \big[\frac{K_{k(\beta_i)}
(\beta_i)}{K_0(\beta_i)}\big]^{|\beta_i|}
\ee
for $k=0$ (and, in particular, for any loop in the the plane), and it
vanishes identically otherwise. We can now take the continuum limit.
The result is simply:
\be \label{4.5}
\chi(\alpha) =  \lim_{a\to 0}\chi^{a,L_x,L_\tau} (\alpha)
=\exp[-\frac{g_0^2}{2}\sum_{i=1}^n k(\beta_i)^2 {\rm Ar}(\beta_i)].
\ee
if $k=0$, and $\chi(\alpha) = 0$ if $k\not=0$, where ${\rm Ar}(\beta)$
is the area enclosed by the loop $\beta$.

A number of remarks are in order. i) We see explicitly the area law in
(\ref{4.5}) (which would signal confinement in $d+1>2$ dimensions).
The same is true for non-Abelian groups. ii) As in the case of the
$\lambda\phi^4$-model in 2-dimensions, no renormalization of the bare
coupling $g_0$ was necessary in order to obtain a well-defined limit.
This is a peculiarity of 2 dimensions. Indeed, in higher dimensions
because the bare coupling does not even have the correct physical
dimensions to allow for an area law. iii) It is interesting to note
that the above expression is completely insensitive to the fact that
we have not taken the infinite volume limit, $L_x,L_\tau\to\infty$ on
the plane, and $L_\tau\to\infty$ on the cylinder. (The only
requirement so far is that the lattice is large enough for the loop
under consideration to fit in it.) Thus, the task of taking these
``thermodynamic'' limits is quite straightforward. iv) In higher
dimensions, one can formulate a program for constuction of the measure
along similar lines (although other avenues also exist). This may be
regarded as a method of obtaining the continuum limit in the lattice
formulation. The advantage is that if the limit with appropriate
properties exists, one would obtain not only the Euclidean
``expectation values'' of Wilson loop functionals but also a genuine
measure on $\agb$, a Hilbert sapce of states and a Hamiltonian.

Finally, let us compare our result with those in the literature.
First, there is complete agreement with \cite{GKK}. However our method
of calculating the vacuum expectation values, $\chi(\alpha)$, of the
Wilson loop observables is somewhat simpler and more direct.
Furthermore, it does not require gauge fixing or the introduction of a
vector space structure on $\ag$ and we were able to treat the cases
$M=\R^2$ and $M=S^1\times\R$ simultaneously.  More importantly, we
were able to obtain a closed expression in the $U(1)$ case. Finally,
the invariance of $\chi(\alpha)$ under area preserving diffeomorphisms
is manifest in this approach; the huge symmetry group of the classical
theory transcends to the quantum level. This important feature was not
so transparent in the previous rigorous treatments (except for the
second paper in \cite{GKK}).

\subsection{A proposal for Constructive Quantum Gauge Field theory}

In this section, we will indicate how one might be able to arrive at a
rigorous, non-perturbative formulation of quantum gauge theories in
the continuum.  The idea of course is to define a {\it quantum theory
of gauge fields} to be a measure $d\bar{\mu}$ on $\agb$ --whose
support may be considerably smaller than $\agb$-- satisfying the
analogs of the Osterwalder-Schrader axioms.  These must be adapted to
the kinematical non-linearity of gauge theories. We will discuss how
the key axioms can be so formulated. As one would expect, they are
satisfied in the 2-dimensional example discussed above. Using them, we
will arrive at the Hamiltonian framework which turns out to be
equivalent to that obtained from canonical quantization.

The two key axioms in the kinematically linear case were the Euclidean
invariance and the reflection positivity. These can be extended as
follows. Let $E: \; \R^{d+1}\to \R^{d+1}$ denote an Euclidean
transformation.  We require:
\be
\chi(E\circ\alpha)=\chi(\alpha)
\ee
for all $E$ in the Euclidean group.  The axiom of reflection
positivity also admits a simple extension. Let us construct the
linear space ${\cal R}^+$ of complex-valued functionals
$\Psi_{\{z_i\},\{\beta_i\}}$ on $\agb$ of the form:
\be
\Psi_{\{z_i\},\{\beta_i\}} (\bar{A}) =
\sum_{i=1}^n z_i\ \bar{T}_{\beta_i}[\bar{A}] =
\sum_{i=1}^n z_i\ \bar{A}[\beta_i] \ ,
\ee
where $\beta_i\in \cal HG$ are independent in the sense of the previous
subsection and have support
in the positive half space, and $z_i$ are arbitrary complex numbers.
Then a measure $d\bar\mu$ on $\agb$ will be said to satisfy the
reflection positivity axiom if:
\be (\Psi,\Psi):=<\theta\Psi,\Psi>:=\int_{\agb}
(\theta\Psi(\bar{A}))^\star\Psi(\bar{A}) \ d\bar{\mu}(\bar{A})\ge 0\; ,
\ee
where $\theta$, as before, is the ``time-reflection'' operation
\footnote{This formulation is for the case when the gauge group is
$U(1),\mbox{ or }SU(2)$. For $SU(N)\; N>2$, the product of traces of
holonomies is not expressible as a linear combination of traces of
holonomies.  Therefore, the argument of the functionals $\chi$ contain
$1, 2, ..., N-1$ loops. However, the required extension is
immediate.}.
The remaining Osterwalder-Schrader axioms can be extended to gauge
theories in a similar manner \cite{S, ALMMT}, although a definitve
formulation is yet to emerge.

Let us now consider the 2-dimensional example for $G=U(1)$. Since the
generating functional (\ref{4.5}) is invariant under all area
preserving diffeomorphisms, it is, in particular, invariant under the
2-dimensional Euclidean group on the plane
\footnote{On the cylinder the Euclidean group, of course, has to be
replaced by the isometry group of the metric, that is, the group
generated by the Killing vectors $\partial_\tau\mbox{ and } \partial_x$.}
The verification of reflection positivity requires more work. We will
simultaneously carry out this verification {\it and} the construction
of the physical Hilbert space. Let us apply the Osterwalder-Schrader
algorithm, outlined in section 2.2, to construct the physical Hilbert
space and the Hamiltonian. First, a careful analysis provides us the
null space ${\cal N}$: its elements are functionals on $\agb$ of the
type:
\be
\tilde{\Psi}[\bar{A}] = \big(\sum_{i=1}^n z_i\ \bar{T}_{\beta_i}
[\bar{A}]\big)-\big(\sum_{i=1}^n z_i
\chi(\beta_i)\big)\ \bar{T}_p[\bar{A}]\ ,
\ee
provided all $\beta_i$ are homotopically trivial. The physical Hilbert
space ${\cal H}$ is given by the quotient construction, ${\cal H} =
{\cal R}^+/\cal N$. On the plane, it is simply the linear span of the
trivial vector $T_p=1$; ${\cal H}$ is one-dimensional.

Let us now consider the more interesting case of the cylinder. If one
of $\beta_i$ or $\beta_j$ contains a homotopically trivial loop
and the other does not, $(\theta\beta_i)^{-1}\circ\beta_j$ necessarily
contains a homotopically non-trivial loop and the characteristic
function (\ref{4.5}) vanishes at that loop. Finally, if both loops are
homotopically non-trivial, then $(\theta\beta_i)^{-1}\circ\beta_j$
necessarily contains a homotopically non-trivial loop unless the
effective winding numbers of the non-trivial loops in $\beta_i$ and
$\beta_j$ are equal. Therefore, the linear spans of $T_\alpha$'s with
zero and non-zero $k$ are orthogonal under $(.,.)$. The former was
shown to coincide with the null space generated by $T_p$.
To display the structure of the Hilbert space ${\cal H} = {\cal R}^+/
{\cal N}$, let us introduce a horizontal loop $\gamma$ at $\tau = 0$
with winding number one.  Then, any loop with winding number $k=n$ can
be written as the composition of $\gamma^n$ and homotopically trivial
loops.  Therefore $\beta=\gamma^{-n}\circ\alpha$ has zero winding
number.  Finally, using $\theta\gamma=\gamma$, it is easy to show that
$T_{\gamma^n\circ\beta}-\chi(\beta) T_{\gamma^n}\in\cal N$, implies
that the Hilbert space ${\cal H}$ is the completion of the linear span
of the vectors $T_{\gamma^n}\; n\in \Z$.  The vectors
$\psi_n:=T_{\gamma^n}$ form an orthonormal basis in $\cal H$. Since
the quotient of ${\cal R}^+$ by the null sub-space ${\cal N}$ has a
positive definite inner-product, $(.,.)$, it is clear that the
generating function $\chi$ satisfies reflection positivity.  Finally,
note that, since the loop $\gamma$ probes the generalized connections
$\bar{A}$ only at ``time'' zero, the final result is analogous to that
for a scalar field where the Hilbert space constuction can also be
reduced to to the fields at ``time'' zero.

Our next task is to construct the Hamiltonian $H$. By definition, $H$
is the generator of the Euclidean time translation semi-group. Let
$\gamma(\tau):=T(\tau)\gamma$ be the horizontal loop at time $\tau$
and set $\alpha=\gamma\circ\rho\circ\gamma(\tau)^{-1}\circ\rho^{-1}$,
where $\rho$ is the vertical path between the vertices of the
horizontal loops.  Then, we have $T_{\gamma(\tau)^n}
=\chi(\alpha^n)T_{\gamma^n}$ as elements of ${\cal H}$, so that
\ba
(\psi_m,\ T(\tau)\psi_n)&=&\chi(\alpha^n)\delta_{n,m}=
[\exp (-\textstyle{1\over 2} n^2 g_0^2 L_x \tau)]\ \delta_{n,m}\\
&=:& (\psi_n,\exp(-\tau H)\psi_m)\ .
\ea
Finally, the completeness of $\{\psi_n\}_{n\in \Z}$ enables us to
write down the action of the Hamiltonian $H$ simply as:
\be
H \psi_n=\frac{g_0^2}{2}L_x E^2 \psi_n,\; E \psi_n=-i  n \psi_n\ .
\ee
Finally, it is clear that the vacuum vector is unique and given by
$\Omega(\bar{A})=1$. Thus, because the key (generalized) Osterwalder
axioms are satisfied by our continuum measure, we can construct the
complete Hamiltonian framework.

To conclude, we note that the 2-dimensional model can also be
quantized directly using the Hamiltonian methods \cite{SS, ALMMT}. The
resulting quantum theory is completely equivalent to the one obtained
above, starting from the Euclidean theory.

\section{Discussion}

In this contribution we first pointed out that, in its standard form
\cite{GJ}, the basic framework of constructive quantum field theory
depends rather heavily on the assumption that the space of histories
is linear. Since this assumption is not satisfied in gauge theories
(for $d+1>2$), a fully satisfactory treatment of quantum gauge fields
would require an extension of the framework. We then suggested an
avenue towards this goal.

The basic idea was to regard $\ag$ as the classical space of histories
and to attempt to construct a quantum theory by suitably completing it
and introducing an appropriate measure on this completion.
To achieve this, one can exploit the ``non-linear duality''
between loops and connections. More precisely, if one uses loops as
probes --the counterpart of test functions in the case of a scalar
field-- one can follow the general methods used in the kinematically
linear theories and introduce the notion of cylindrical functions and
cylindrical measures on $\ag$. The question is if these can be
extended to genuine measures. The answer turned out to be in the
affirmative: there exists a completion $\agb$ of $\ag$ such that every
cylindrical measure on $\ag$ can be extended to a regular,
$\sigma$-additive measure on $\agb$. Thus, we have a ``non-linear
arena'' for quantum gauge theories; the discussion is no longer tied
to the linear space of tempered distributions.  We were able to
indicate how the Osterwalder-Schrader axioms can be generalized to
measures on the non-linear space $\agb$. The key open problem is that
of singling out {\it physically appropriate} measures.

The space $\agb$ is analogous to the algebraic dual $\shb$ of the
Schwarz space one encounters in the kinematically linear case.
Therefore, it is almost certainly too big for quantum Yang-Mills
theories (although there are indications that it is of the ``correct
size'' for diffeomorphism invariant theories such as general
relativity.) That is, although measures which would be physically
relevant for Yang-Mills theories could be well-defined on $\agb$,
their support is likely to be significantly smaller. In the
kinematically linear case, the Bochner-Milnos theorem provides tools
to find physically relevant measures and tells us that their support
is the space $S'$ of tempered distributions. The analogous result is,
unfortunately, still lacking in our extension to gauge theories.
Without such a result, it is not possible to specify the exact
mathematical nature of the Schwinger functions of the theory --the
``Euclidean expectation values'' of the Wilson loop operators.  This
in turn means that we have no results on the analytic continuation of
these functions, i.e., on the existence of Wightman functions. What
we {\it can} formulate, is the notion of reflection positivity and
this ensures that the physical Hilbert space, the Hamiltonian and the
vacuum exists.

It is clear from the above discussion that our framework is
incomplete. We need to introduce appropriately weak topologies on the
space of hoops --the probe space-- and find generating functionals
$\chi(\alpha)$ which are continuous with respect to them.  Only then
can one have sufficient control on the nature of Schwinger functions.
For the moment, $\agb$ serves only as the ``universal home'' for the
measures we want to explore. As we saw, this strategy was successful
in the 2-dimensional Yang-Mills theory.

In higher dimensions, there are reasons to be concerned that $\agb$
may be too large to play even this ``mild'' role. That is, one might
worry that the elements of $\agb$ are allowed to be so
``pathological'' that it would be difficult to define on them the
standard operations that one needs in mathematical physics. For
instance, $\agb$ arises only as a (compact, Hausdorff) topological
space and does not carry a manifold structure. For physical
applications on the other hand, one generally needs to equip the
domain spaces of quantum states with operations from differential
calculus. Would one not be stuck if one has so little structure? It
turns out that the answer is in the negative: Using projective
techniques associated with families of graphs, one {\it can} develop
differential geometry on $\agb$. In particular, notions such as vector
fields, differential forms, Laplacians and heat kernels are well
defined \cite{AL2}. Thus, at least for the purposes we want to use
$\agb$, there is no {\it obvious} difficulty with the fact that it is
so large.

A more subtle problem is the following. A working hypothesis of the
entire framework is that the Wilson loop operators should be
well-defined in quantum theory. From a ``raw,'' physical point of
view, this would seem to be a natural assumption: after all Wilson
loop functionals are the natural gauge invariant observables. However,
technically, the assumption {\it is} strong. For example, if the
connection is assumed to be distributional, the Wilson loop
functionals cease to be well-defined. Therefore, in a quantum theory
based on such a hypothesis, the connection would not be representable
as an operator valued distribution. In particular, in the case of a
Maxwell field, such a quantum theory can not recover the textbook Fock
representation \cite{AI2}. More generally, the representations that
can arise will be {\it qualitatively} different from the Fock
representation in which the elementary excitations will not be plane
waves or photon-like states. Rather, they would be ``loopy,''
concentrated along ``flux lines.'' They would be related more closely
to lattice gauge theories than to the standard perturbation theory.
In 2 dimensions, this reprentation does contain physically relevant
states.  Whether this continues to be the case in higher dimensions is
not yet clear. It is conceivable that the quantum theories that arise
from our framework are only of mathematical interest. However, even if
this turns out to be the case, they would still be of considerable
significance since as of now there does not exist a single quantum
gauge theory in higher dimensions. Finally, if it should turn out that
loops are too singular for physical purposes, one might be able to use
the extended loop group of Gambini and co-workers \cite{G} as the
space of probes.  This group has the same ``flavor'' as the hoop group
in that it is also well tailored to incorporate the kinematical
non-linearities of gauge theories. However, the hoops are replaced by
extended, smoothened objects so that Fock-like excitations are
permissible.

To conclude, our main objective here is to revive intertest in
manifestly gauge invariant approaches to quantum gauge theories, in
which the kinematical non-linearities are met head on right from the
beginning.  There have been attempts along these lines in the past
(see, particularly \cite{S}) which, however, seem to have been
abandoned.  (Indeed, to our knowledge, none of the major programs for
construction of quantum Yang-Mills theories is still being actively
pursued.)  The specific methods we proposed here are rather tentative
and our framework is incomplete in several respects. Its main merit is
that it serves to illustrate the type of avenues that are available
but have remained unexplored.

\end{document}